# Decameter Type IV Burst Associated with a behind-the-limb CME Observed on 7 November 2013

**V.N. Melnik[1], A.I. Brazhenko[2], A.A. Konovalenko[1], V.V. Dorovskyy[1], H.O. Rucker[3],M. Panchenko[4], A.V. Frantsuzenko[2], M.V. Shevchuk [1]**

**Abstract**
We report on the results of observations of a type IV burst by URAN- 2 (*Ukrainian Radio interferometer of Academy Sciences*) in the frequency range 22 - 33 MHz, which is associated with the CME (coronal mass ejection) initiated by a behind-the-limb active region (N05E151). This burst was observed also by the radio telescope NDA (*Nancay Decameter Array*) in the frequency band 30–60 MHz. The purpose of the article is the determination of the source of this type IV burst. After analysis of the observational data obtained with the URAN-2, NDA, STEREO (*Solar-Terrestrial Relations Observatory*) A and B spacecraft, and SOHO (*Solar and Heliospheric Observatory*) spacecraft we come to the conclusion that it is a core of a behind-the-limb CME. We conclude that the radio emission can escape the center of the CME core at a frequency of 60 MHz and originates from the periphery of the core at frequency 30 MHz due to occultation by the solar corona at corresponding frequencies. We find plasma densities in these regions supposing the plasma mechanism of radio emission. We show that the frequency drift of the start of the type IV burst is governed by an expansion of the CME core. Type III bursts, which were observed against this type IV burst, are shown to be generated by fast electrons propagating through the CME core plasma. A type II burst registered at frequencies 44 – 64 MHz and 3 – 16 MHz was radiated by a shock with a velocity of about 1000 km $s^{-1}$ and 800 km $s^{-1}$, respectively.

V.N. Melnik
melnik@rian.kharkov.ua

[1] Institute of Radio Astronomy, National Academy of Sciences of Ukraine, Kharkov, Ukraine

[2] Gravimetrical Observatory, Poltava, Ukraine

[3] Commission for Astronomy, Austrian Academy of Sciences, Graz, Austria

[4] Space Research Institute, Austrian Academy of Sciences, Graz, Austria

## 1. Introduction

Type IV bursts were classified as a separate type of solar radio phenomenon by Boischot (1957). They distinguish stationary and moving type IV bursts respectively. For a long time type IV bursts were observed only in the decimeter and meter ranges (Stewart, 1985). Coronal arches and CMEs are considered to be responsible for the former and latter, respectively. Regular observations of these bursts at frequencies 8-32 MHz were started during Cycle 23 with the radio telescopes UTR-2 (*Ukrainian T-shaped Radio telescope of two modification*) and URAN-2 (*Ukrainian Radio interferometer of Academy Scienses*) due to the broadband 60-channel spectrometers and DSP (*digital spectral polarimeter*) (Melnik, Rucker, and Konovalenko, 2008). With the launch of WIND and later STEREO (*Solar-Terrestrial Relations Observatory*) A and B spacecraft high quality observations of type IV bursts in the hectometer range became possible (Gopalswamy, 2011; Hillaris, Bouratzis, and Nindos, 2016).

There are some issues concerning the mechanism of type IV bursts radio emission, the locations from which radio emissions at different frequencies escape, beaming of radiation of these bursts, their associations with type II bursts and SEP (Solar Energetic Particle) events, and others (Stewart, 1985; Aschwanden, 2004; Gopalswamy, 2011), which remain as open questions.

A recent article (Gopalswamy *et al.*, 2016) considered the event on 7 November 2013. At this day, a CME (coronal mass ejection) was ejected from a behind the limb region (N05E151). It was accompanied by a type IV burst observed by STEREO A and B from 10:40 to 11:50 UT and was not registered by WIND. The authors concluded that the directivity of the type IV radio emission was narrow, about $60^{\circ}$, along its propagation path and that the observed type IV burst was stationary and not moving.

This day the radio telescope URAN-2 (Poltava, Ukraine) carried out observations from 06:00 to 13:30 UT in the frequency band 17 - 33 MHz. Starting from 10:22 UT and up to 10:44 UT this type IV burst was registered by this radio telescope. Approximately at the same time this burst was also observed by the *Nancay Decameter Array* (NDA, Nancay, France) in the frequency band 30 - 60 MHz.

Thus, apparently there is a discrepancy between the results of cosmic and ground-based observations. It can be solved when the location of the observed type IV burst is determined.

The purpose of the present article is to answer the question what is the source of the radio emission of the type IV burst observed on 7 November 2013 at frequencies 22 - 60 MHz, based on the observations of the URAN-2, NDA, STEREO A and B, SOHO. The type III bursts and type II burst which are associated with this type IV burst and regions of their emissions are discussed as well.

## 2. Observations

The radio telescope URAN-2 is situated near Poltava (Ukraine) (Brazhenko *et al.*,

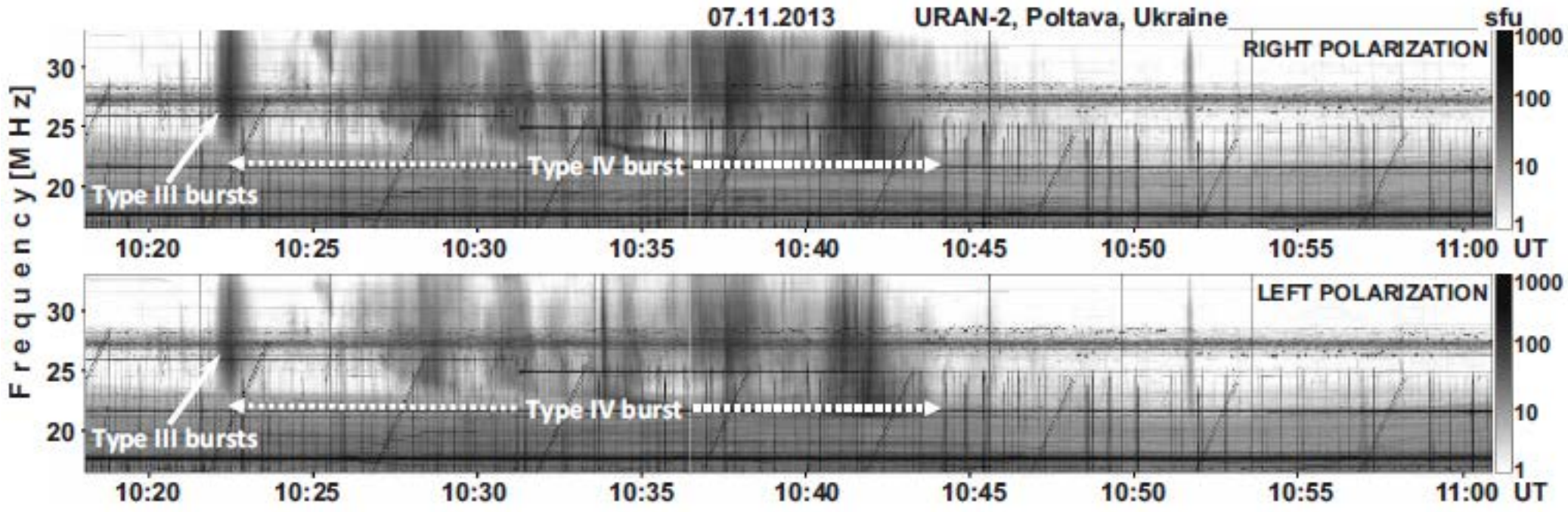

**Figure 1**. Type IV burst and type III bursts on 7 November 2013, according to URAN-2.

2005). Its effective area is about 28,000 square metres that provides very high sensitivity of observations. The sensitivity of the URAN-2 has been calibrated using the galactic background radiation as a reference radio source at different frequencies and different directions. The working frequency band is 8 - 33 MHz but this day observations were carried out from 17 to 33 MHz because of ionosphere interferences. Registration was performed by DSPz (Ryabov *et al.*, 2010), with the frequency and time resolutions of 4 kHz and 100 ms respectively. The radio telescope also measured the degree of circular polarization of the registered radio emission.

In Figure 1 the dynamic spectrum of solar radio emission from 10:17 to 11:01 UT in the frequency band 17 - 33 MHz is shown. It is seen that the type IV burst began at approximately 10:22 UT and finished at 10:44 UT. There were also some type III bursts against the type IV burst.

The fluxes of the type IV burst and type III bursts reached 30 - 40 sfu and 100 sfu respectively (Figure 2). Polarizations of these bursts were high, slightly below 40% for the type IV burst and up to 50% for the type III bursts. This supports the idea that both type IV burst and type III bursts are generated at the first harmonic of the local plasma frequency (Stewart, 1985). Type III bursts were rather diffuse and consisted of sub-bursts with positive and negative frequency drift rates of 2 - 6 MHz $s^{-1}$ (Figure 3). These sub-bursts can not be the effects of ionosphere influence because this day we registered ionosphere disturbances only up to 22 MHz while these sub-bursts were observed at frequencies higher than 22 MHz. Note that similar sub-bursts are usual phenomenon for decameter type IV bursts (Melnik, Rucker, and Konovalenko, 2008; Melnik *et al.*, 2010; Antonov *et al.*, 2014). The drift rate of the start of the type IV burst is about 30 kHz $s^{-1}$. Indeed the leading edge of the type IV burst appeared at frequency $f_1$ = 33 MHz at approximately $t_1$ = 10:22 UT, and at frequency $f_2$ = 22 MHz at $t_2$ = 10:30 UT, that gives the frequency drift rate $\left|(f_2 - f_1)/(t_2 - t_1)\right| \approx 30 kHz/s$.

This day the radio telescope NDA observed the Sun as well (Figure 4). Its data demonstrate exactly the same set of bursts, which were recorded by the URAN-2 instrument but at frequencies 30 - 60 MHz. A weak type II burst with a drift rate of about 250 kHz $s^{-1}$ was observed at frequencies 64 - 44 MHz from 10:16 to 10:17 UT. A burst with such a drift rate should appear at a frequency 32 MHz at approximately 10:18 UT. And indeed, the weak burst with flux and polarization of about 10 sfu and 20 %, respectively, is observed at this frequency and approximately at this time (Figure 2).

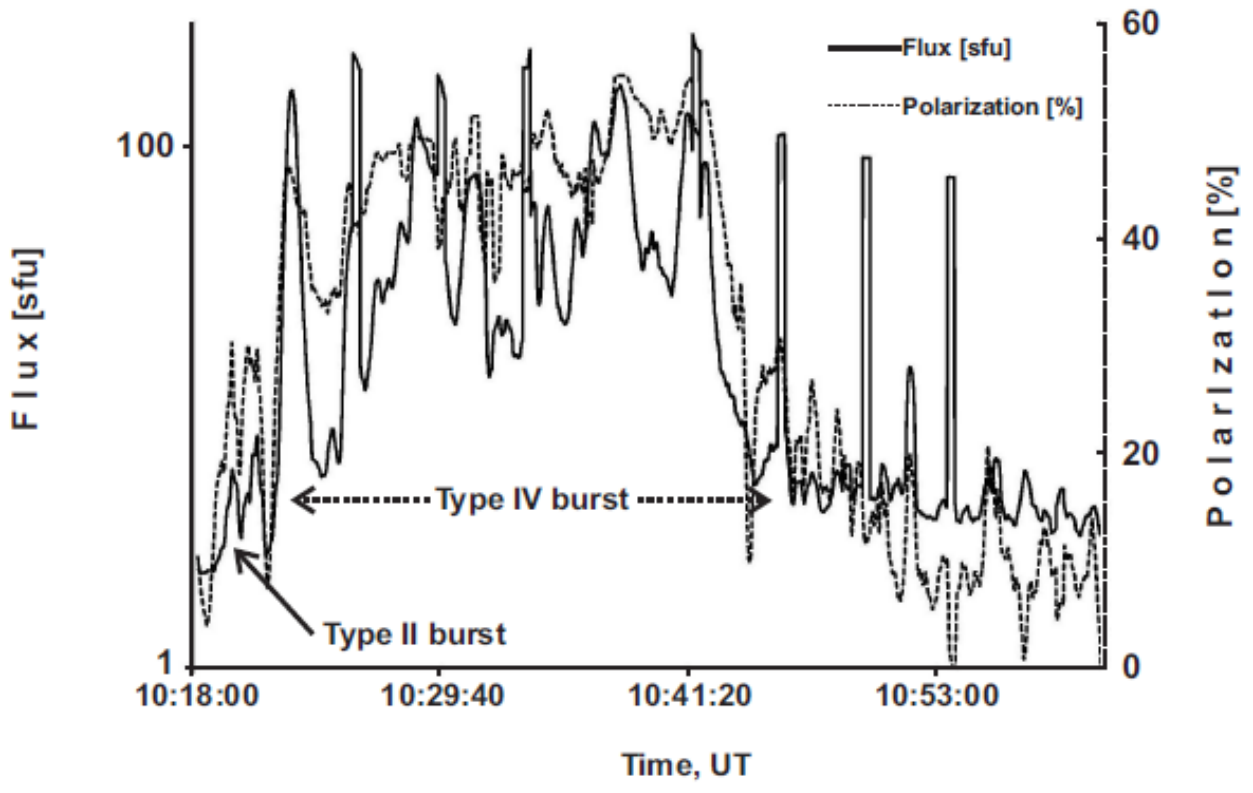

**Figure 2.** The flux and polarization profiles of the type IV burst and type III bursts at 32 MHz.

We also attempted to find higher frequency data from Phoenix, CALLISTO (*Compoud Astronomical Low cost Low frequency Instrument for Spectroscopy and Transportable Observatory*), and ARTEMIS (*Appareil de Routine pour le Traitement et l'Enregistrement Magnetique de l' Information Spectral* ) instruments but failed. The radio telescopes, which were involved in the CALLISTO project, carried out the observations of the solar radio emission this day, but none of them registered the type IV burst. Apparently their sensitivity is not enough to register this event. At the same time CALLISTO registered bright type III bursts at 10:22 and 10:36 UT, which was also observed by URAN-2, NDA and STEREO A and B.

As reported by Gopalswamy *et al.* (2016) STEREO A and B registered the type IV burst, type III bursts and type II burst approximately at the same time (Figure 5) as the radio telescopes URAN-2 and NDA but shortly later. Gopalswamy *et al*, (2016) associated this event with a CME, which was initiated by an active region at the longitude E151° that appeared to be behind the limb as seen from the Earth. This CME propagated practically in the direction towards STEREO B (Figure 6d) so it was registered by STEREO B as a halo-CME. At the same time STEREO A registered this CME as a compact object with bright core and envelope (Figure 6a). The velocity of the CME nose was high, up to 2000 km $s^{-1}$ (Gopalswamy *et al.*, 2016). At the same time we derived that the velocity of the CME core was only 550 km $s^{-1}$ according to the STEREO A and SOHO data. STEREO B observed the type IV burst very clearly (Figure 5) but it was very weak in the records of STEREO A (Figure 5). The cause of this effect is that the viewing was high (about 60°) for STEREO A, but it is important that despite higher viewing angles ground-based radio telescopes could register this type IV burst due to their high sensitivity (Konovalenko et al., 2013). According to STEREO B the type IV burst was observed mainly between 16 to 11 MHz from 10:50 to 11:30 UT. The fact that at low frequencies the burst was registered later than at the higher frequencies (Figure 7) points out that this type IV burst was moving. This is confirmed by the drift of forehand not only at frequencies 22 - 33 MHz but also at frequencies 11 - 16 MHz (Figure 7). Its drift rate is about 6 kHz $s^{-1}$ at frequencies 11 - 16 MHz.

Figure 7 shows type III bursts having their counterparts at higher frequencies, but there are also type III bursts at 10:18 UT and about 10:58 UT, which are absent in the URAN-2 and DA records. Both STEREO A and B observed a type II burst, which drifted from 16 MHz to 3 MHz with a drift rate of about 8 kHz $s^{-1}$(Figure 5).

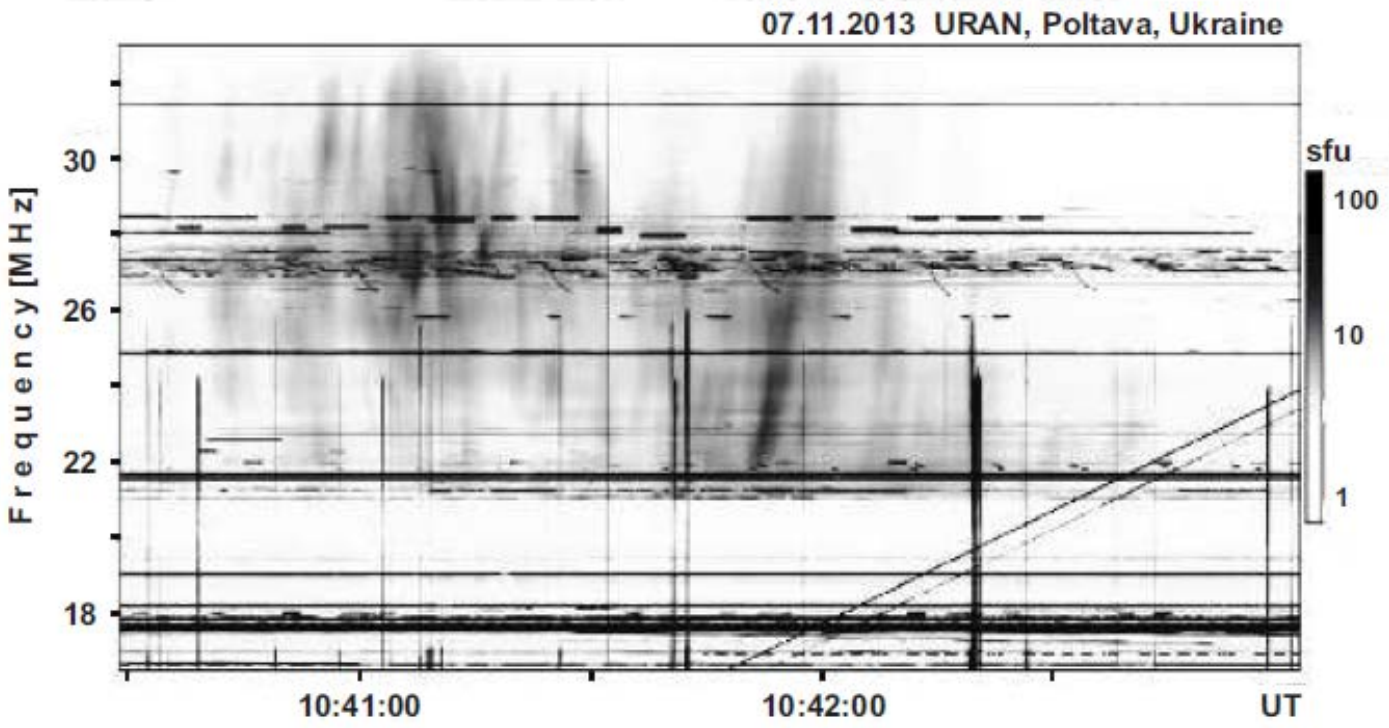

**Figure 3.** Type III bursts with fine structure in the form of sub-bursts with positive and negative drift rates.

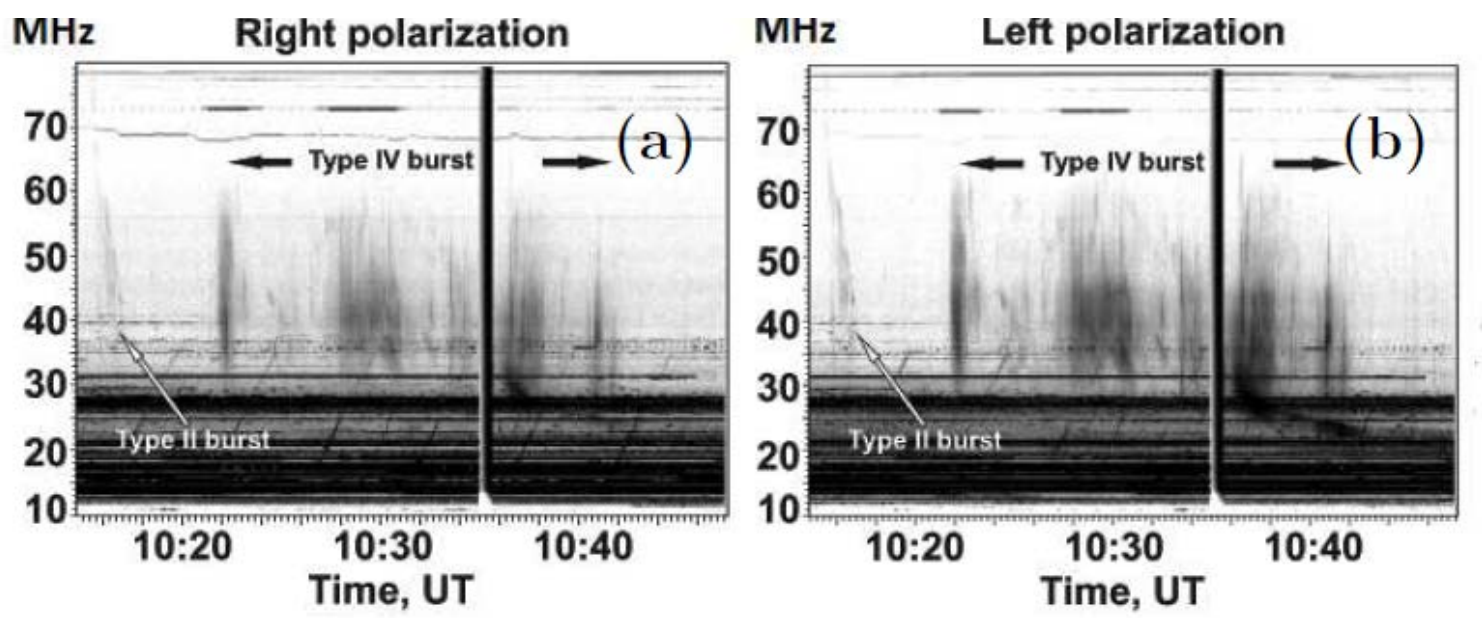

**Figure 4**. Type IV burst, type III bursts and type II burst observed by the radio telescope NDA on 7 November 2013.

## 3. Discussion

The fact that both the type IV burst and type III bursts associated with a behind-the-limb CME are seen at frequencies 22 - 60 MHz with ground-based radio telescopes, raises the question about source locations of these bursts and their association with the CME. The locations and sizes of the CME core (obtained from STEREO A data) at 10:22 UT and 10:28 UT, when the type IV burst was observed in the frequency ranges 30 - 60 MHz and 23 - 60 MHz, are shown in Figure 8a. Also the levels of local plasma frequencies 23, 30 and 60 MHz and the regions of occultation by the Newkirk coronal density model (Newkirk, 1961) (on this particular day there were active regions on the eastern limb) are given. From Figure 8a we see that radio emissions at 23, 30 and 60 MHz can only escape towards Earth from the regions to the left from the lines $f_{NK}$ = 23, 30, 60 MHz, respectively. If we suppose that the source of the type IV burst is the CME core (following Bain et al, (2014)) then it is reasonable to suggest that radio emissions at frequencies 23 and 30 MHz escape from peripheral regions of the core and at frequency 60 MHz from its center. Assuming plasma mechanism of radio emission generation, we find densities in the center ($n_c = 4.5 \times 10^7 cm^{-3}$) and in peripheral regions ($n_p = 1.3 \times 10^7 cm^{-3}$) of the CME core. If the density distribution in the core is exponential

$$n(r) = n_c \exp(-\alpha r) \quad (1)$$

here $r$ is the distance from the core center and $\alpha = 2.5 / R_\odot$ ($R_\odot$ is the solar radius) taking into account that at $r = 0.5R_\odot$ (the core radius at 10:22 UT) and the density equals $n(r = 0.5R_\odot) = n_p = 1.3 \times 10^7 cm^{-3}$. For compensation of gas kinetic pressure p = nkT at core surface by the magnetic pressure $p_B = B^2 / 8\pi$ the magnetic field [B] of $6 \times 10^{-2} G$ at plasma temperature in the core $T = 10^5 K$ and $0.6G$ at temperature $T = 10^6 K$ are required. The density of the surrounding coronal plasma at distance $R = 2.8R_\odot$ equals only $n_{cp} = 1.2 \times 10^6 cm^{-3}$ which is essentially smaller than the density in the core at this distance.

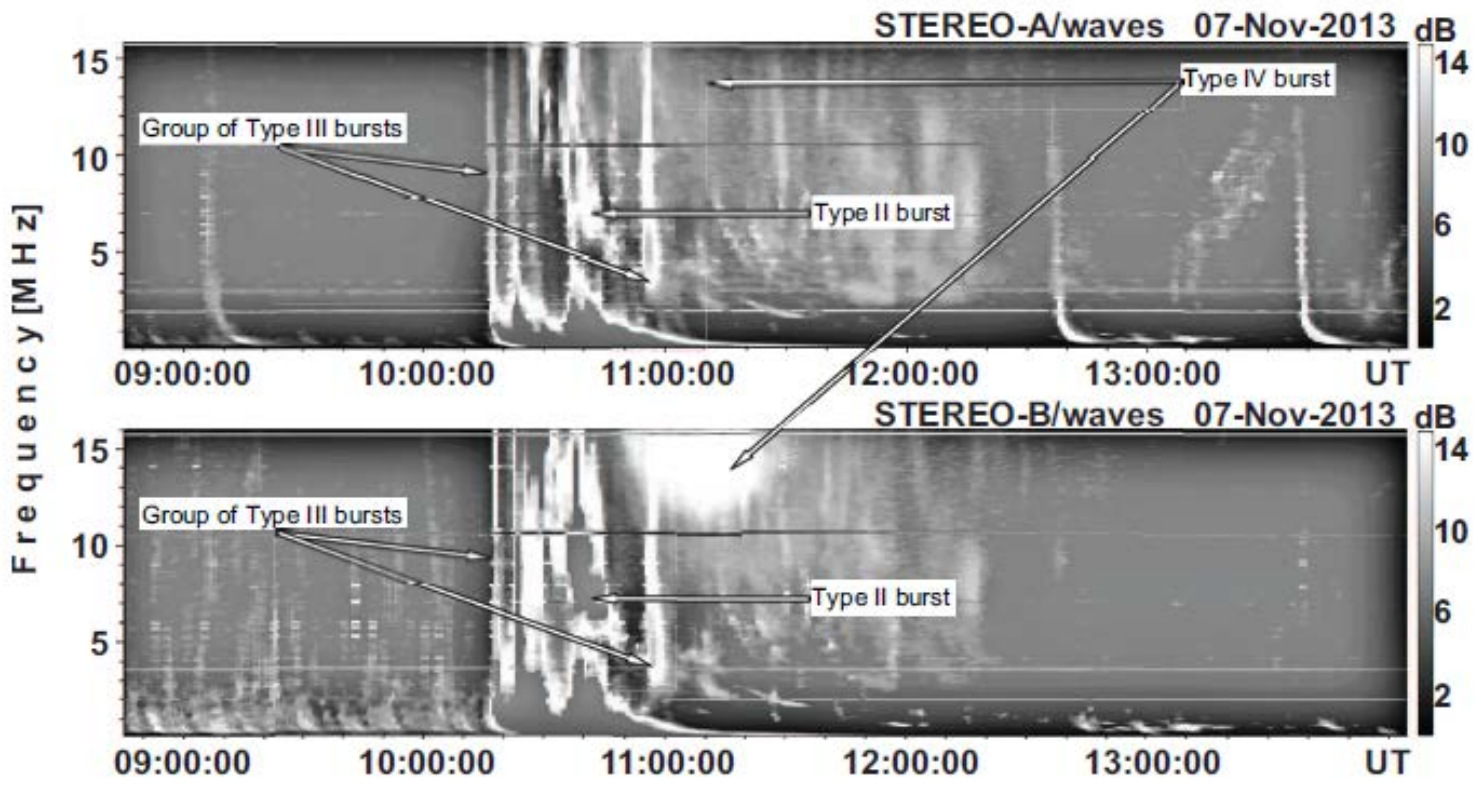

**Figure 5**. Type IV burst, type III bursts and type II burst observed by STEREO A and B.

Supposing that the size of the type IV source was about 1 solar radius at frequencies 22 - 30 MHz we derived the brightness temperature of the type IV radio emission of $T_{br} = 10^9 K$. Such a high brightness temperature is difficult to explain by gyrosynchrotron mechanism (Stewart, 1985) but is easily interpreted within the framework of the plasma mechanism (Stewart, 1985). Following the density distribution Equation 1 the mass of the CME core can be estimated as

$$M_{CME} = 4\pi \int_0^{R_\odot/2} m_p n(r) r^2 dr \approx 10^{16} g \quad (2)$$

This value is in agreement with the SOHO data (https://cdaw.gsfc.nasa.gov). Due to the expansion of the core the periphery density decreases and this leads to the frequency drift of the type IV burst whose drift rate is defined by the equation

$$\frac{df}{dt} = \frac{f}{2n} \frac{dn}{dr} \frac{dr}{dt} \quad (3)$$

where $dr / dt$ is the velocity of the core expansion which is equal to 300 km $s^{-1}$ according to the STEREO A data. Taking into account the inhomogeneity size ($(dn / ndr)^{-1}$ we derive the drift rate from the Equation 3, which is about 27 kHz $s^{-1}$ in the frequency range 22 - 33 MHz, that is in good agreement with the results of observations by URAN-2.

As we see in Figures 1, 4, 5, and 7 the separate type III bursts and their groups at 10:22 UT, 10:29 UT, 10:32 UT, 10:35 UT, 10:37 UT and 10:42 UT were registered by the NDA, URAN-2, STEREO A, and STEREO B. But it is worthwhile to notice that if type III bursts at frequencies 22 - 60 MHz were observed against the type IV burst, these bursts below 16 MHz appeared before the type IV burst. At the same time rather powerful type III bursts at 10:18 UT and 10:58 UT observed by STEREO A and B (Figure 5, 7) were absent on the dynamic spectrum of the URAN-2 (Figure 1, 7) and NDA (Figure 4). Such specific feature of appearance of type III bursts can be understood in the following way: bursts visible by all radio telescopes leave the CME core, whilst they are radiated at frequencies 22 - 60 MHz and out of the core, when they are radiated at frequencies below 16 MHz. If these bursts generated the radio emission only at frequencies 22 - 60 MHz from outside the core (the coronal plasma) the ground based radio telescopes could not register them because of occultation by the solar corona (Figure 8a). Consequently, we conclude that we observed fast electrons which propagated through plasma of the CME core along helical magnetic field of flux rope (Aschwanden, 2004) at first and after that these electrons leave the core and propagated in the coronal plasma ahead of the CME core. Moreover, from Figure 5 we see that these electrons pass the front of the shock, which is responsible for type II burst recorded by STEREO A and B. The fact that radio emission at frequencies 22 - 60 MHz escapes the CME core is confirmed by the diffuse character of type III bursts at these frequencies and by a large number of sub-bursts with different positive and negative drift rates (Figure 3). This reflects the complex structure of core plasma with different inhomogeneities.

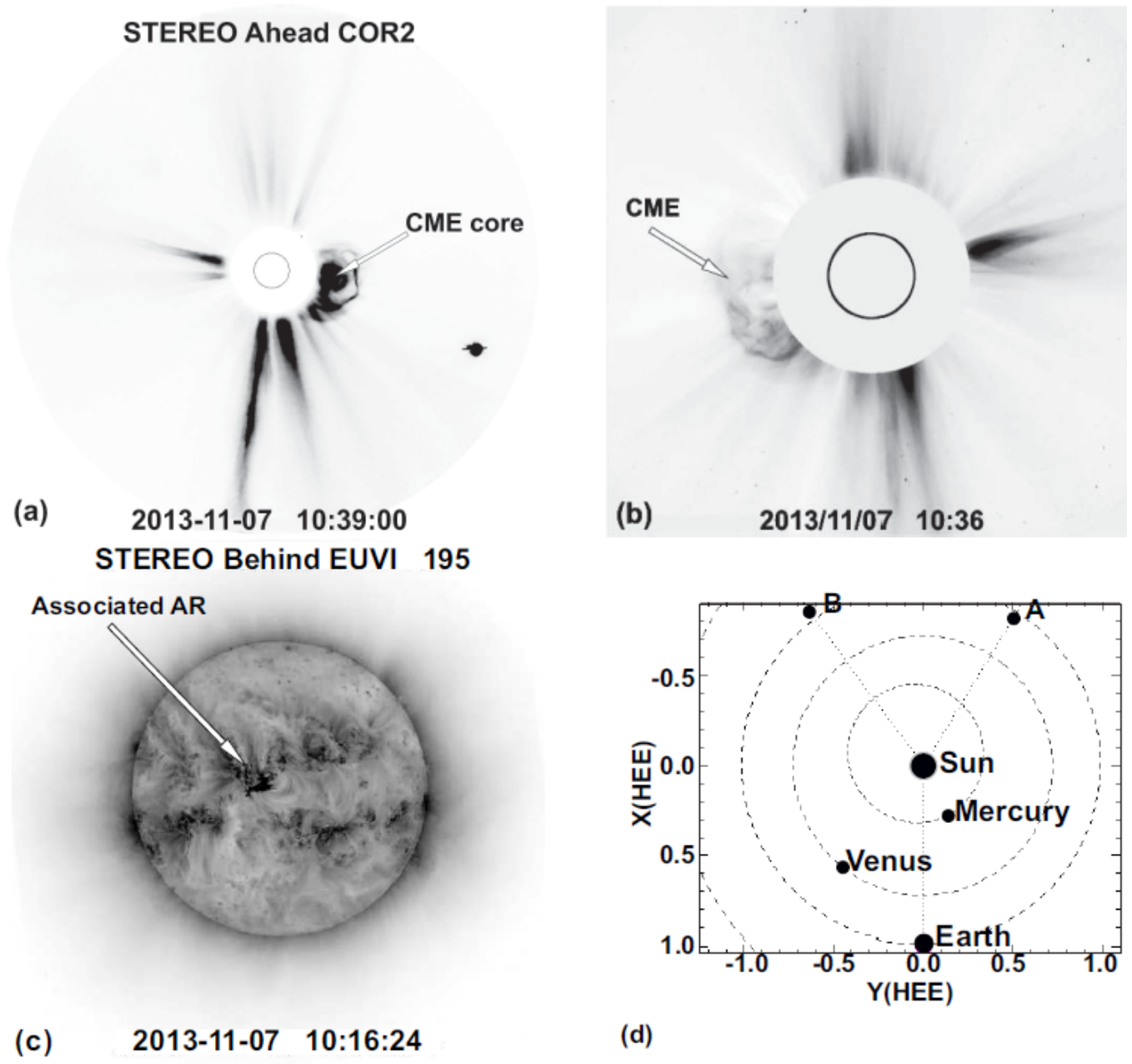

**Figure 6**. CME according to STEREO A (a), and SOHO (b) and the active region responsible for the CME (c) and STEREO A and B on the Earth orbit on 7 November, 2013 (d).

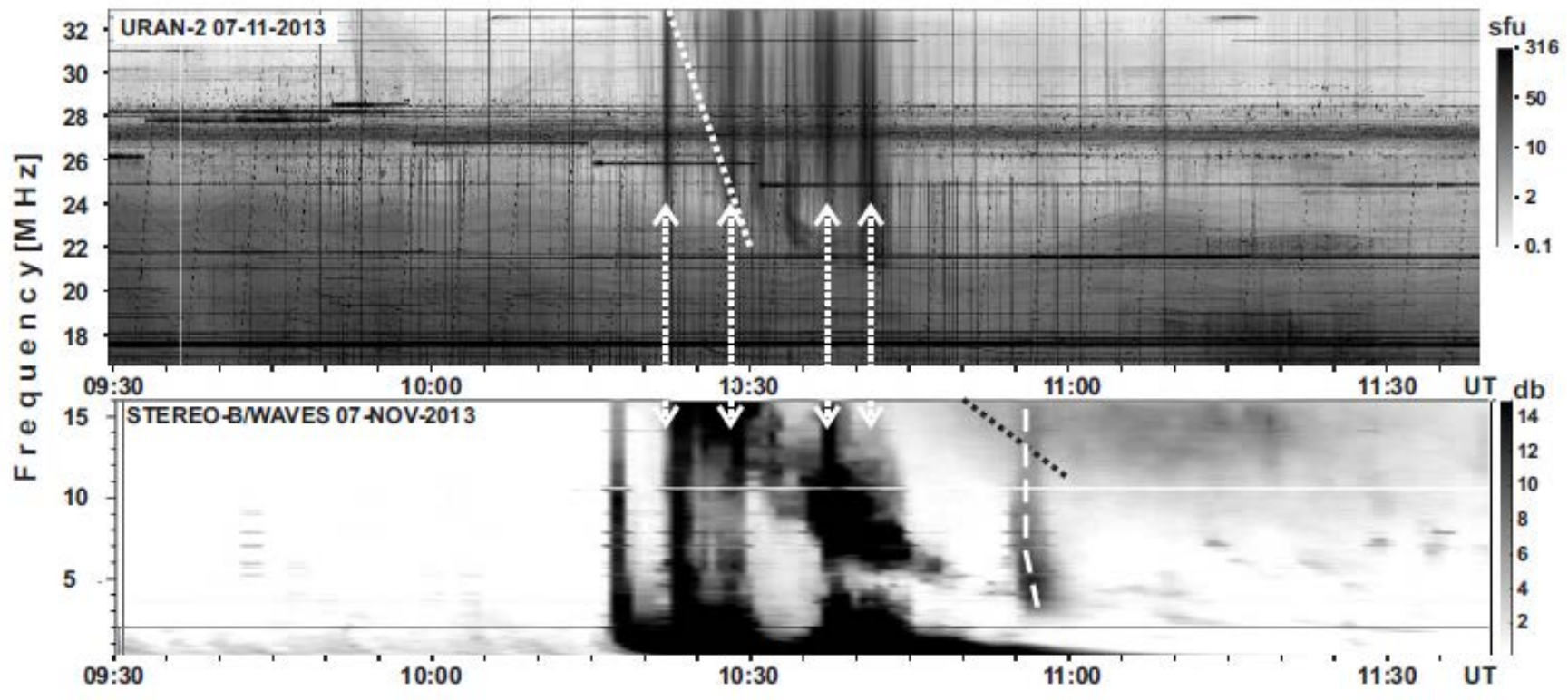

**Figure 7**. Moving type IV burst and type III bursts according to URAN-2 and STEREO B data on 7 November 2013. The drift of type IV burst is shown by dotted lines and the drift of type III burst (10:48 UT) by the dashed line.

The type III burst at 10:18 UT (Figure 7) escaped from the coronal plasma ahead of the CME and as a result the ground-based radio telescopes could not observe this burst due to occultation, while STEREO A and B registered it. Similar situation took place at the observation of unusual behind-limb decameter burst on 3 June 2011 (Melnik et al., 2014). Analogically the type III burst at 10:58 UT (Figure 7) could not be observed by the radio telescopes URAN-2 and NDA because the type III electrons reached the distances 1.4 - 2R_ in the solar corona, from which the radio emissions at frequencies 22 - 60 MHz originated, 30 minutes after the passage of the CME. This type III burst shows a dog-leg of dynamic spectrum (Figure 7), which can be interpreted as a transition of type III electrons from CME core into interplanetary space.

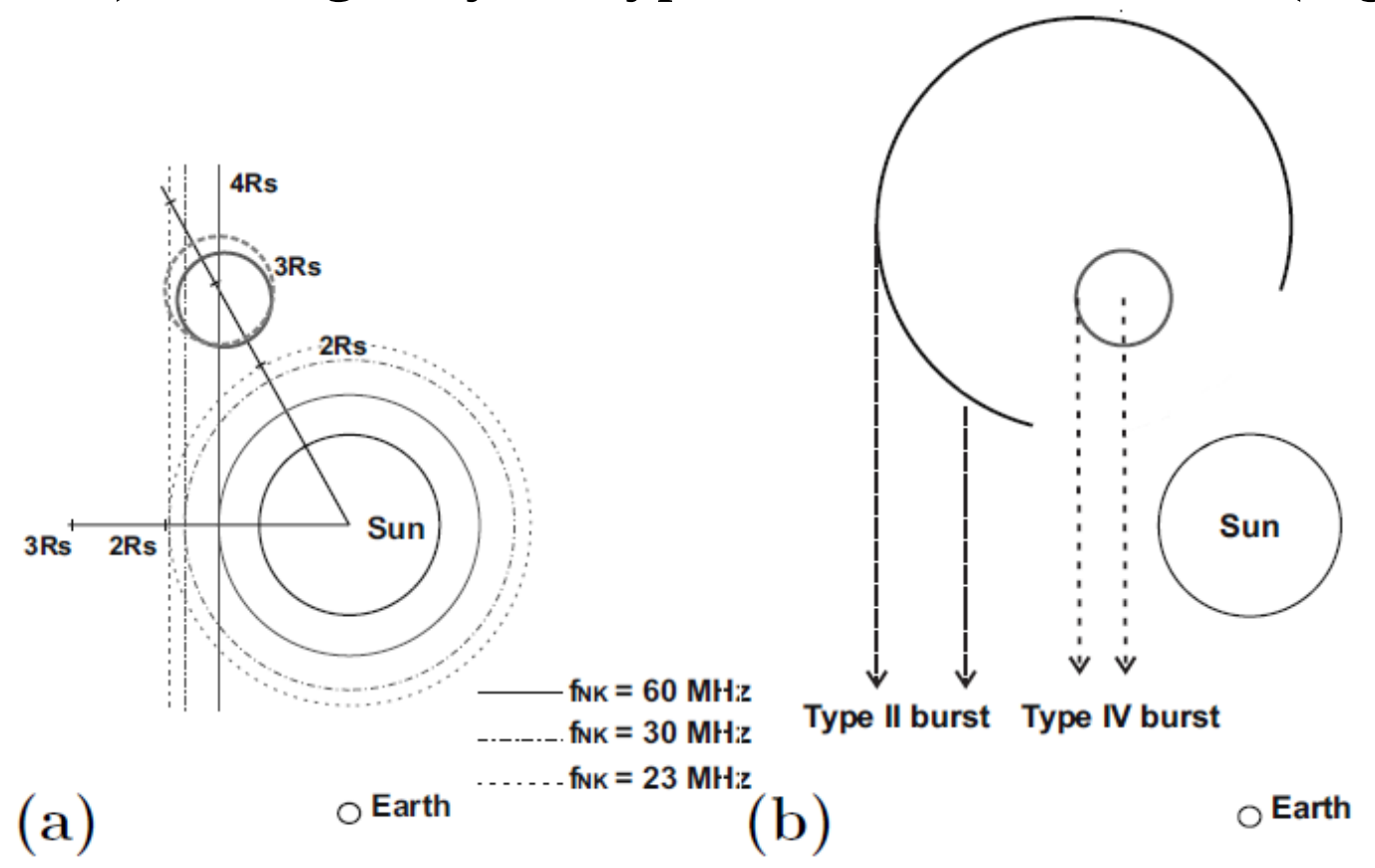

**Figure 8**. (a)Schematic positions of the CME core at 10:22 UT (solid) and 10:28 UT (dashed), the Sun and the Earth and levels in the solar corona, which correspond to local plasma frequencies 23, 30 and 60 MHz. (b) The type II burst and type IV burst escape the leading edge of the CME and the CME core, respectively.

The abovementioned explanation regards to the case if discussed type III bursts were initiated by the active region (N05E151). Of course, some type III bursts could be associated with other active regions including situated on the visible disk and then these bursts would not be connected with the discussed CME. But at this time namely the region (N05E151) was active and we suggested the current explanation as the most possible candidate.

It is well-known (Nelson and Melrose, 1985) that shocks are responsible for type II bursts. In our case this is driven-CME shock, which coincided with the leading edge of CME (see, for example Gopalswamy, 2011) (Figure 8b). The frequency drift rate of the type II burst is derived as the longitudinal velocity $V_{II} = V\cos\theta$ ($V$ is the velocity of the leading edge and $\theta$ is the angle between the radial direction and the direction of source motion) from the equation (Gopalswamy, 2011)

$$\frac{1}{f}\frac{df}{dt} = \frac{V_{II}}{2H} \quad (4)$$

where $f$ f is the frequency of observation, $df\ /\ dt$ is the frequency drift rate of the type II burst, $H = (dn / ndr)^{-1}$ is the density scale height of the coronal plasma. The velocity $V_{II}$ equals 1000 km s$^{-1}$ in the frequency range 44 - 64 MHz and is slightly smaller, 800 km s$^{-1}$, in the frequency range 3 - 16 MHz. Taking into account that the leading nose of CME had a velocity of about 2000 km s$^{-1}$ and did not essentially depend upon $\theta$ we concluded that the observed type II bursts were generated by the flank parts of the shock ($\theta = 60^{\circ} - 70^{\circ}$). Such a situation is standard for type II and IV events (see, for example, Gopalswamy, 2011; Dorovskyy et al., 2015). Note, that Figure 8b shows directions of the type II and type IV radio emissions towards the Earth. Certainly, the emissions of these bursts propagate in the directions to STEREO A and B as well. Moreover, it is naturally to consider that di_erent parts of the spherical shock driven by the CME can be sources of the type II bursts. Thus, instruments with different locations can register the radio emission as from the same or different parts of the spherical shock.

**Conclusion**

We report on the results of observations of type IV burst associated with a behind-the-limb CME (E150_) by ground-based radio telescopes URAN-2 and NDA in the frequency range 22 - 60 MHz. This burst can be attributed to moving type IV bursts with a frequency drift rate of about 30 kHz s$^{-1}$ at frequencies 22 - 33 MHz. We come to the conclusion that the type IV burst is the moving one and the source of its emission in the frequency band 22 - 60 MHz is the core of the CME, which propagates in the behind-the-limb region. Also we determine the density distribution in the core based on the plasma emission mechanism of the type IV burst. Thus in the core center the density equals to $n_c = 4.5 \times 10^7 cm^{-3}$, and at the periphery $n_p = 1.3 \times 10^7 cm^{-3}$.

We also determine that the type III bursts, which are visible against the type IV burst are radiated by the electrons which passed through the CME core. The type II burst, observe at frequencies 44 - 64 MHz and 3 - 16 MHz, is the radio emission of the ank shock wave caused by the CME.

**Acknowledgments** We thank the referee for useful remarks and suggestions, which improved the manuscript.

**Disclosure of Potential Conflicts of Interest** The authors declare that they have no conflicts of interest.